\documentclass{article}
\usepackage{spconf,amsmath,graphicx, amssymb}
\usepackage{float}
\usepackage{cite}
\usepackage{mathtools}

\usepackage{capt-of}
\usepackage{url}
\usepackage[nolist]{acronym}
\usepackage{amssymb,multirow}
\usepackage{booktabs, bm}
\usepackage{booktabs,caption,fixltx2e}
\usepackage{subfigure}
\usepackage{lineno}
\usepackage[]{algorithm2e}
\usepackage{hyperref}
\usepackage{bibspacing}
\usepackage{caption}
\def\H{{\mathsf H}}
\def\T{{\mathsf T}}
\def\CC{{\mathbb C}}
\def\RR{{\mathbb R}}

\newacro{MVDR}[MVDR]{Minimum Variance Distortionless Response}
\newacro{GEV}[GEV]{Generalized Eigenvalue Decomposition}
\newacro{MCWF}[MCWF]{Multi-Channel Wiener Filter}
\newacro{SDW-MWF}[SDW-MWF]{Speech Distortion Weighted MWF}
\newacro{SISDR}[SI-SDR]{Scale Invariant Signal-to-Distortion Ratio}
\newacro{STFT}[STFT]{Short-Time Fourier Transform}
\newacro{DNN}[DNN]{Deep Neural Network}
\newacro{SNR}[SNR]{Signal-to-Noise-Ratio}
\newacro{FLOPs}[FLOPs]{Floating Point Operations}
\newacro{SIR}[SIR]{Signal-to-Interference Ratio}
\newacro{ASR}[ASR]{Automatic Speech Recognition}
\newacro{IBM}[IBM]{Ideal Binary Mask}
\newacro{IRM}[IRM]{Ideal Ratio Mask}
\newacro{WLM}[WLM]{Wiener-Like Mask}
\newacro{SDR}[SDR]{Signal-to-Distortion Ratio}
\newacro{STOI}[STOI]{Short-time Objective Intelligibility}
\newacro{MSE}[MSE]{Mean-Squared Error}
\newacro{WER}[WER]{Word Error Rate}
\newacro{WPE}[WPE]{Weighted Prediction Error}
\newacro{MISO}[MISO]{Multiple Input Single Output}
\newacro{DFL}[DFL]{Deep Feature Loss}

\title{Towards low-distortion multi-channel speech enhancement:\\ The ESPNet-SE submission to the L3DAS22 challenge}

\name{
\begin{tabular}{c}
\it Yen-Ju Lu$^{1,5}$, Samuele Cornell$^2$, Xuankai Chang$^1$, Wangyou Zhang$^4$, Chenda Li$^4$, \\
Zhaoheng Ni$^3$, Zhong-Qiu Wang$^1$, Shinji Watanabe$^1$
\end{tabular}
}
\address{
$^1$Carnegie Mellon University, USA\,\,
$^2$Università Politecnica delle Marche, Italy\,\,
$^3$Meta AI, USA \\
$^4$Shanghai Jiao Tong University, Shanghai\,\,
$^5$Academia Sinica, Taipei \\
{\small\texttt{\{cornellsamuele, neilyenjulu, nizhaoheng, wang.zhongqiu41\}@gmail.com}}
}

\begin{document}
\ninept
\maketitle

\setlength{\abovedisplayskip}{3.5pt}
\setlength{\belowdisplayskip}{3.5pt}

\begin{abstract}
This paper describes our submission to the L3DAS22 Challenge Task 1, which consists of speech enhancement with 3D Ambisonic microphones. The core of our approach combines Deep Neural Network (DNN) driven complex spectral mapping with linear beamformers such as the multi-frame multi-channel Wiener filter. Our proposed system has two DNNs and a linear beamformer in between. Both DNNs are trained to perform complex spectral mapping, using a combination of waveform and magnitude spectrum losses. The estimated signal from the first DNN is used to drive a linear beamformer, and the beamforming result, together with this enhanced signal, are used as extra inputs for the second DNN which refines the estimation. Then, from this new estimated signal, the linear beamformer and second DNN are run iteratively. The proposed method was ranked first in the challenge, achieving, on the evaluation set, a ranking metric of $0.984$, versus $0.833$ of the challenge baseline. 
\end{abstract}

%\section{Task1 System Description}\label{sec:sys_des}
\section{Introduction}\label{sec:intro}
Multi-channel speech enhancement (SE) is an important pre-processing step for many applications, such as 
hands-free speech communication, hearing aids, smart speakers,
and automatic speech recognition (ASR) \cite{vincent2018audio}.
In its broad definition, it consists of joint denoising and dereverberation of a desired target speech signal from a noisy-reverberant multi-channel mixture signal captured by a microphone array. % signal model in time-domain 
This arduous problem has been effectively addressed in the last decade with DNN-based methods, which have been firmly established as the de-facto mainstream approach for speech enhancement \cite{WDL2018}.
Multi-channel DNN-based methods can be roughly divided into \emph{hybrid} \cite{heymann_blstm, Boeddeker_practical, erdogan2016improved, ochiai_unified_2017, Wang2021Sequential} and \emph{fully-neural} \cite{luo2019fasnet, ren2021neural,  tolooshams2020channel, liu2020multichannel, Wang2021MCCSM} methods. %tolooshams2020channel
The former combines DNNs with conventional signal processing based techniques, using the DNNs to drive, for example, dereverberation algorithms such as \ac{WPE} \cite{Kinoshita2017} or classical beamforming algorithms such as \ac{MVDR} and \ac{MCWF} \cite{heymann2016neural, vincent2018audio}.
%Instead, i
In fully-neural systems, DNNs are trained to directly estimate the target speech from the mixture. The DNN can have either a \ac{MISO} structure or can be used to directly estimate linear beamforming filters in the \ac{STFT} domain \cite{ren2021neural} or in the time domain \cite{luo2019fasnet}.
Fully-neural methods are effective and often outperform hybrid techniques for what regards signal-level SE metrics such as \ac{SISDR} \cite{le2019sdr}, \ac{STOI} \cite{taal2011algorithm} etc.
However, unlike conventional dereverberation and beamforming algorithms, they tend to introduce non-linear distortions, which can degrade the performance of downstream tasks, such as \ac{ASR} \cite{iwamoto2022bad,zhang2021closing}.
This problem can be mitigated by using end-to-end training or \ac{DFL} \cite{bagchi2018spectral, li2021espnet}. On the other hand, these techniques requires re-training or fine-tuning whenever the back-end model changes and are cumbersome to apply when there are multiple downstream tasks. 

This trade-off between signal-level SE metrics and \ac{ASR} performance is at the core of the L3DAS22 speech enhancement challenge since the models are ranked by considering both \ac{STOI} \cite{guizzo2021l3das22} and \ac{WER}. While correlated to some degree, the two metrics reflect two highly different downstream application scenarios: \ac{ASR} and human listening for \ac{STOI}. Our goal is to devise a  ``generalist" SE model, optimized independently from these metrics or the backend \ac{ASR} models, but able to significantly improve both.

To address this arduous problem, we employ a  framework derived from previous works \cite{Wang2021FCP, Wang2021LowDistortion}, which combines the merits of hybrid and fully-neural methods: namely, beamforming's ability at producing low-distortion estimates and DNN's high capacity at suppressing non-target signals.
Compared with \cite{Wang2021FCP}, in this study we perform multi-channel enhancement by beamforming directly on 3D Ambisonic microphone format. 
We introduce here one main novelty: 
applying a multi-frame beamformer \cite{nakatani2019unified, zhang2020multi, Wang2021Sequential} with the beamforming filter estimated directly from the DNN estimated target signal. We show that this helps to tackle the problem of target signal misalignment explained in Section \ref{sec:task_des}.
We propose an iterative neural/beamforming enhancement (iNeuBe) architecture inculding two TCN-DenseUNet \cite{Wang2021LowDistortion} which are employed in a MISO configuration and a beamformer.
Our system is depicted in Fig.~\ref{systemoverview}.
%Both DNNs are used to regress directly the complex \ac{STFT} coefficients of the target signal. 
The first DNN (DNN$_1$) takes in input the complex \ac{STFT} coefficients of the multi-channel input mixture signal ($\mathbf{Y}$) and regresses directly the complex \ac{STFT} coefficients of the target signal ($\hat{S}^{(1)}$).
DNN$_1$ enhanced signal ($\hat{S}^{(1)}$) is used to drive a multi-frame MCWF (mfMCWF) at the first ($i=0$) iteration to derive a low-distortion estimate of the target signal ($\hat{S}^{(\text{mfMCWF})}_{i=0}$). Both $\hat{S}^{(\text{mfMCWF})}_{i=0}$ and $\hat{S}^{(1)}$ are then used as additional features for the second DNN ($\text{DNN}_2$) to refine the target estimate. The output of $\text{DNN}_2$ ($\hat{S}^{(2)}_{i=0}$) can be used iteratively in place of $\hat{S}^{(1)}$ to compute another refined beamforming result $\hat{S}^{(\text{mfMCWF})}_{i=1}$ which is then fed back to  $\text{DNN}_2$ together with $\hat{S}^{(2)}_{i=0}$.

The proposed framework placed first in the L3DAS22 speech enhancement challenge, achieving a Task 1 challenge metric of $0.984$ on the evaluation set, versus $0.833$ achieved by the official baseline and $0.975$ by the runner-up system. This indicates that the proposed approach is a promising step towards ``generalist'' multi-channel SE, as it achieves both low \ac{WER} and high \ac{STOI} without any fine-tuning with the back-end \ac{ASR} model or use of \ac{STOI}-derived losses. 

We 
%make
have made our implementation available through the ESPNet-SE toolkit \cite{li2021espnet}.

\section{L3DAS22 Task 1 Description}\label{sec:task_des}

%L3DAS 2022 description.

%Ambisonic

%Given a $P$-microphone noisy-reverberant mixture, the physical model in the time-domain can be formulated as.

The L3DAS22 3D speech enhancement task (Task 1) \cite{guizzo2021l3das22} challenges participants to predict the dry speech source signal from its far-field mixture recorded by two four-channel Ambisonic-format signals in a noisy-reverberant office environment.
The challenge dataset is ``semi-synthetic''.
It consists of 252 measured room impulse responses (RIRs).
The dry speech source signals are sampled from LibriSpeech \cite{panayotov2015librispeech} and the dry noise signals from FSD50K \cite{fonseca2020fsd50k}. 
Two first-order A-format Ambisonic microphones arrays, each with four microphones, are employed to record the RIRs. A single room is used for RIR measurement. The microphone placement is fixed, with one at the room center and the other 20\,cm apart.
Notably, the room configuration and microphone placement do not change between training and testing, and the source positions are distributed uniformly inside the room. 
Artificial mixtures are generated by convolving dry speech and noise signals with the measured RIRs and mixing the convolved signals together.
The \ac{SNR} is sampled from the range $[6, 16]$ dBFS (decibels relative to full scale).
The generated A-format Ambisonic mixtures are then converted to B-format Ambisonic. %via a transformation consisting of a pre-filter, a mixing matrix, and a post-filter. 
The total amount of data is around 80 hours for training and 6 hours for development. 

The challenge ranks the submitted systems using a combination of \ac{STOI} \cite{taal2010short} and \ac{WER}:
\begin{equation}\label{}
     \text{Task1 Metric} = \big(\text{STOI} + (1 - \text{WER}^{\bot 1})\big)/2.
\end{equation}
where $\text{WER}^{\bot 1} = \text{min}(\text{WER}, 1)$. %, preventing to increase the score by simply introducing insertions. 
%challenge measures the  scores of each submitted system, and ranks their performance according to the following composite metric:
The values of STOI and WER are both in the range of $[0,1]$, so is the composite metric.
%The WER is calculated by the speech recognition results for the enhanced speech through a pre-trained Wav2Vec2 model \cite{baevski2020wav2vec}.
%Differently from the evaluation setup commonly employed for ASR applications, 
The WER is computed based on the transcription of the estimated target signal and that of the reference signal, both decoded by a pre-trained Wav2Vec2 ASR model \cite{baevski2020wav2vec}.

%Note that differently from many other setup, in this task, the goal is to predict the dry source signal from the far-field multi-channel mixture. Thus performing at the same time dereverberation and suppression of the interfering signal. 
We emphasize that the goal of the challenge is recovering the dry speech signal from a far-field noisy-reverberant mixture.
As such, the metrics above are computed with respect to the dry ground truth. %signal.
This makes the task extremely challenging, because, besides removing reverberation and noises, a system also needs to time-align the estimated signal with the dry speech signal in order to obtain a good \ac{STOI}. 
\ac{STOI}, in fact, contrary to \ac{WER}, is highly sensitive to time-shifts: e.g. a shift in the order of 100 samples alone can decrease the \ac{STOI} value from $1.0$ to $0.9$ for the very same oracle target speech. Thus it is required that the model performs, either implicitly or explicitly, localization of the target source inside the room, so that an aligned estimate can be produced. 

\section{Proposed Method}\label{sec:sys_des}

\subsection{System Overview}

Let us denote the dry speech source signal as $s[n]\in \RR$ and the far-field mixture recorded by Ambisonic microphones as $\mathbf{y}[n]\in \RR^{P}$, where $n$ indexes discrete time and $P$ ($=8$ in this study) is the number of channels.
Following the challenge baseline \cite{ren2021neural}, our proposed system operates 
%in the STFT domain and directly on the B-format Ambisonic microphones signals ($2$ microphones with $4$ channels each).
on the STFT spectra of the B-format Ambisonic signals.
We denote the STFT coefficients of the mixture and dry speech signal at time $t$ and frequency $f$ as $\mathbf{Y}(t,f)\in \CC^{P}$ and $S(t,f)\in \CC$, respectively.
For simplicity, we will omit in the following the $t$ and $f$ indexes, and denote the STFT spectra simply as $\mathbf{Y}$ and $S$, and signals as $\mathbf{y}$ and $s$.

\begin{figure}
  \centering  
  %\captionsetup{justification=centering}
  \includegraphics[width=8.5cm]{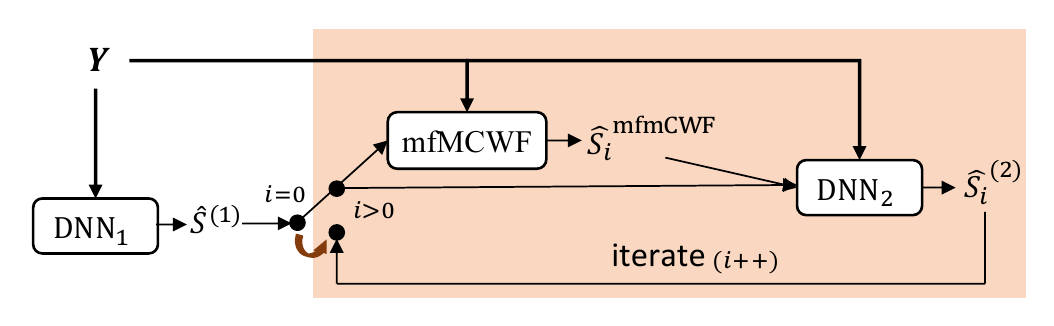} \vspace{-0.2cm} \\  
  \caption{Overview of proposed iterative neural/beamforming enhancement (iNeuBe) framework. A multi-frame multi-channel Wiener filter (mfMCWF) beamformer is applied between the two DNN MISO networks.} %with mfMCWF.}
  \label{systemoverview}%\vspace{-0.4cm}
  \vspace{-0.5cm}
\end{figure}

Our proposed iNeuBe framework, illustrated in Fig.~\ref{systemoverview}, contains two DNNs and a linear beamforming module in between.
Both DNNs have a MISO structure and are trained using multi-microphone complex spectral mapping \cite{Wang2020MCCSMDereverb, Wang2021MCCSM, Tan2022}, where the real and imaginary (RI) components of multiple input signals are concatenated as input features for the DNNs to predict the RI components of the target speech.
%We can also include the RI components of the beamforming results as the input features.
%As input features they employ the real and imaginary (RI) components of the multi-channel input mixture STFT representation $\bm{Y}$, and output the RI components of the estimated target anechoic signal STFT representation $S$.

More in detail, for DNN$_1$ we concatenate the RI components of $\mathbf{Y}$ as input to predict the RI components of $S$.
DNN$_1$ produces an estimated target speech $\hat{S}^{(1)}$, which is at the first iteration $i=0$ used to compute an mfMCWF for the target speech.
Subsequently, the RI components of the beamforming result $\hat{S}^{\text{mfMCWF}}_{i=0}$, the input mixture $\mathbf{Y}$, and $\hat{S}^{(1)}$ are concatenated %along the channel axis
and fed as input for DNN$_2$ to further refine the estimation
for the RI components of $S$.
DNN$_2$ produces another refined estimation of $S$, i.e. $\hat{S}^{(2)}_{i=0}$, which can be used iteratively in place of $\hat{S}^{(1)}$ to recompute the beamforming result and as an additional feature to DNN$_2$. 

%we can refine the beamformer and run a second pass of DNN$_2$ to further improve the target estimation.

%As we outline in Section \ref{sec:results} using just DNN$_1$ output already gives substantial improvements over the challenge baseline system and also a conventional DNN-mask based beamformer based on ConvTasNet \cite{luo2018convtasnet}. The beamforming module, DNN$_2$ and run-time iterative estimation (up to 2 iterations) further increase performance at the cost of increased computational requirements. 
The rest of this section describes the DNN architecture, the loss function employed for the DNN training, the mfMCWF beamforming algorithm, and the run-time iterative procedure.

%In our experiments, we find that directly predicting the dry source signal based on the far-field mixture already works pretty well.
%Intuitively, this is possible because the microphones are arranged in a 3D, fixed geometry and placed in a fixed acoustic environment, through supervised learning the DNNs could learn to estimate the target location and implicitly compensate the signal shift from cues related to the particular acoustic environment adopted in the challenge. 

\vspace{-0.2cm}
\subsection{Multi-Microphone Complex Spectral Mapping}\label{sec:mc_csm}

We employ the TCN-DenseUNet architecture described in the Fig.~15 of \cite{Wang2021LowDistortion} for both DNN$_1$ and DNN$_2$ (the parameters are not shared).
%complex spectral mapping.
It is a temporal convolution network (TCN) sandwiched by a U-Net derived structure. 
DenseNet blocks are inserted at multiple frequency scales of the encoder and decoder of the U-Net.
This network architecture has shown strong performance in tasks such as speech enhancement, dereverberation and speaker separation \cite{Wang2021FCP, Wang2021MCCSM, Wang2021LowDistortion}.
The network takes as input a real-valued tensor with shape $C\times T \times F$, where $C$ is the number of channels, $T$ the number of STFT frames and $F$ the number of STFT frequencies.
The RI components of different input signals are concatenated along the channel axis and fed as feature maps to the network.
In the case of DNN$_1$, $C$ equals 16 as we have 8 microphone channels.
Linear activation units are used in the output layer to obtain the predicted RI components of the target signal. 
Each network has around 6.9 million parameters.

Given the DNN-estimated RI components, denoted as $\hat{R}^{(b)}$ and $\hat{I}^{(b)}$ where $b \in \{1,2 \} $ indicates which of the two DNNs produces the outputs, we compute the enhanced speech as $\hat{S}^{(b)}=\hat{R}^{(b)}+j\hat{I}^{(b)}$, where $j$ is the imaginary unit, and use inverse STFT (iSTFT) to re-synthesize the time-domain signal $\hat{s}^{(b)}=\text{iSTFT}(\hat{S}^{(b)})$.
After that, we equalize the gains of the estimated and true source signals by using a scaling factor $\ddot{\alpha}$, and define the loss function on the scaled, re-synthesized signal and its STFT magnitude: 
\begin{align}\label{wav+mag}
\mathcal{L}^{(b)}_{\text{Wav+Mag}} = & 
\| \ddot{\alpha} \hat{s}^{(b)} - s \|_1 + \Big\| |\text{STFT}(\ddot{\alpha}\hat{s}^{(b)})| - |\text{STFT}(s)| \Big\|_1,
\end{align}
where $\| \cdot \|_1$ calculates the $L_1$ norm, $|\cdot|$ computes magnitude, and $\text{STFT}(\cdot)$ extracts a complex spectrogram.
$\ddot{\alpha}={{\text{argmin}}}_{\alpha}\,\| \alpha \hat{s}^{(b)} - s \|_2^2=\big(s^{\T} \hat{s}^{(b)}\big)/\big({\hat{s}^{{(b)}^{\T}}}\hat{s}^{(b)}\big)$, where $\| \cdot \|_2$ computes the $L_2$ norm.
The loss on magnitude can improve metrics such as STOI and WER which favor signals with a good estimated magnitude \cite{Wang2021compensation}.

Before training, we normalize the sample variance of the multi-channel input mixture to $1.0$. We do the same normalization also for the dry speech source signal.
We found this normalization procedure essential for training our models as there is a significant gain mismatch between the reference and the mixture signals. % (near vs. far-field).

\subsection{Multi-Frame MCWF}\label{sec:mfmcwf}

Based on the estimated target signal $\hat{S}^{(b)}$ produced by DNN$_1$ or DNN$_2$, following \cite{Wang2021LowDistortion} we compute an mfMCWF per frequency through the following minimization problem:
\begin{align}\label{mfmcwf}
\underset{\mathbf{w}(f)}{{\text{min}}} 
\sum_t \big|
\hat{S}^{(b)}(t,f) - \mathbf{w}(f)^{\H} \widetilde{\mathbf{Y}}(t,f)
\big|^2,
\end{align}
where $\widetilde{\mathbf{Y}}(t,f)=[\mathbf{Y}(t-l,f)^\T,\dots,\mathbf{Y}(t,f)^\T,\dots,\mathbf{Y}(t+r,f)^\T]^\T$ and $\mathbf{w}(f)\in \CC^{(l+1+r)P}$.
$l$ and $r$ control the context of frames for beamforming, leading to a single-frame MCWF when $l$ and $r$ are zeros, and an mfMCWF when $l$ and $r$ are positive.
%The $\hat{S}^{(b)}(t,f)$ is the output of DNNs, where $b \in \{1,2 \} $ denotes the output of which DNN.
The minimization problem is quadratic, and a closed-form solution $\hat{\mathbf{w}}(f)$ is available:
\begin{align}\label{mcwfcov}
\hat{\mathbf{w}}(f) &= \hat{\mathbf{\Phi}}(f)^{-1} \hat{\mathbf{z}}^{(b)}(f) \\
\hat{\mathbf{\Phi}}(f) &= \sum_t \widetilde{\mathbf{Y}}(t,f) \widetilde{\mathbf{Y}}(t,f)^{\H} \\
\hat{\mathbf{z}}^{(b)}(f) &= \sum_t \widetilde{\mathbf{Y}}(t,f) \hat{S}^{(b)}(t,f)^{*},
\end{align}
where $(\cdot)^{*}$ computes complex conjugate.
The beamforming result $\hat{S}^{\text{mfMCWF}}$ is computed  as:
\begin{equation}\label{}
 \hat{S}^{\text{mfMCWF}}(t,f) = \hat{\mathbf{w}}(f)^{\H} \widetilde{\mathbf{Y}}(t,f).
\end{equation}
Notice that in the computation of $\hat{\mathbf{z}}^{(b)}(f)$ and $\hat{\mathbf{\Phi}}(f)$, we average over all the frames in each utterance and compute a time-invariant beamformer, implicitly assuming that the transfer functions between the arrays and sources do not change within each utterance.
%in the time-frame over which the averaging is performed.
This is a valid assumption for the L3DAS22 setup \cite{guizzo2021l3das22}.
%This is certainly true in our case because of the ``semi-synthetic'' L3DAS22 dataset construction \cite{guizzo2021l3das22}.
We emphasize that our approach directly performs beamforming on Ambisonic signals.

As outlined in Section~\ref{sec:task_des}, the dry source signal is not time-aligned with the far-field mixture.
In this scenario, a multi-frame beamformer is highly desirable, as a larger context of frames can be leveraged by the beamformer to compensate the signal shift.
This DNN-supported mfMCWF was proposed recently in \cite{Wang2021Sequential}.
The major difference is that here we use multi-microphone complex spectral mapping to obtain $\hat{S}^{(b)}$, which consists of DNN-estimated magnitude and phase.
In contrast, \cite{Wang2021Sequential} uses monaural real-valued magnitude masking on the far-field mixture to obtain $\hat{S}^{(b)}$ and hence $\hat{S}^{(b)}$ has the mixture phase. 
When target speech is not time-aligned with the mixture, our approach is clearly more principled, as the DNN is free to estimate an $\hat{S}^{(b)}$ that is sample-aligned with $S$. Instead, if real-valued masking is used, the estimated signal would be aligned with the mixture. 
%, and our experiments on the L3DAS22 dataset seem to support this.

For similar reasons, other multi-frame filters \cite{nakatani2019unified, zhang2020multi} cannot align their predictions with the dry target signal.
In addition, although they have shown good performance for signals recorded by omnidirectional microphones, it is unclear whether they can be directly applied for signals in Ambisonic format.
In contrast, our mfMCWF can readily deal with both formats, without any changes.

%Based on the beamforming result $\hat{S}^{\text{mfMCWF}}$, the second MISO $\text{DNN}_2$ is trained to predict the clean speech $S$ from the concatenation of original input $\mathbf{Y}$, previous enhancement result $\hat{S}^{(1)}$, and beamforming result $\hat{S}^{\text{mfMCWF}}$. 
% The losses for the system are calculated as the difference between the target speech and two estimated speech signals from the two MISO DNNs.

% \begin{equation}
% \mathcal{L}^{(b)} = \parallel |\hat{S}^{(b)}| - |\hat{S}|  \parallel_1
%   \label{}
% \end{equation}

\subsection{Run-Time Iterative Processing}\label{sec:iterative}

Ar run time, we can iterate the orange block in Fig.~\ref{systemoverview} to gradually refine the target estimate. 
Denoting as $\hat{S}^{(2)}_{i=0}$ the output of DNN$_2$ after the first pass, we can use this estimate in place of $\hat{S}^{(1)}$ to run again the beamforming module and obtain  $\hat{S}^{\text{mfMCWF}}_{i=1}$. This new beamformed estimate, together with $\hat{S}^{(2)}_{i=0}$ and $\mathbf{Y}$, can then be fed back again to DNN$_2$ to produce an estimate at iteration two, $\hat{S}^{(2)}_{i=1}$, and so on.

%The final estimated result $\hat{S}^{(3)}$ is from the second pass of DNN$_2$ taking $\mathbf{Y}$, $\hat{S}_q^{(2)}$, and $\hat{S}^{\text{mfMCWF(2)}}$ as inputs.

\section{Experimental Setup}\label{sec:setup}

\subsection{Configurations of Proposed Method}\label{sec:dnnarchitecture}

Regarding our iNeuBe architecture, we use an STFT window size of 32 ms and a hop size of 8 ms.
As analysis window we employ square-root Hanning.
DNN$_1$ and DNN$_2$ are trained separately and in a sequential manner: after DNN$_1$ is trained, we run it on the entire training set to generate the beamforming results and $\hat{S}^{(1)}$, and then train DNN$_2$ using the generated signals as the extra input.
Regarding the mfMCWF module, we compare the performance of setting $l$ and $r$ to different values in Section \ref{sec:results}. 
Regarding the challenge results, we use the immediate outputs from the model without any post-processing.

\subsection{Benchmark Systems}\label{sec:benchmark}

%In Section \ref{sec:results} we will compare our proposed method with other state-of-the-art multi-channel speech enhancement models that we have considered during the challenge participation.
%In particular, we explored the use of 
In this challenge, we also experimented with several state-of-the-art enhancement models such as DCCRN \cite{hu2020dccrn}, Demucs v2 and v3 \cite{defossez2019demucs}, and FasNet \cite{luo2019fasnet}.
We also explored an improved version of the mask-based beamforming model used in \cite{cornell2021fbanks}, which is based on ConvTasNet \cite{luo2018convtasnet}. This model is directly derived from \cite{cornell2021fbanks} and uses ConvTasNet separator to predict a magnitude mask for the target signal. The mask is then used to derive a time-invariant MVDR filter which is employed to estimate the target speech. Differently from \cite{cornell2021fbanks}, here we employ TAC \cite{luo2020end} after every repeat in the ConvTasNet separator (we use the standard values of $R=3$ repeats and $X=8$ blocks). DCCRN, Demucs v2 and v3 systems instead rely on complex spectral mapping, without explicit beamforming operations.

\section{Results}\label{sec:results}

%Table~\ref{tab:dnn1_ablation} reports the challenge metrics obtained on the development set for different DNN$_1$ configurations. 
%We consider different strategies for producing the target estimate.
%In detail we compare complex spectral mapping, which directly predicts target speech, with the strategy adopted by the challenge baseline (also denoted as ``neural beamforming'') \cite{ren2021neural}, where a time-varying beamforming filter is first estimated by a DNN and then applied to the mixture to estimate the target speech.
%In addition, we compare two strategies of computing the mfMCWF beamformer (with $l=4, r=3$), one by using the estimated target speech produced by complex spectral mapping as outlined in Section~\ref{sec:mfmcwf} and another by having DNN$_1$ estimate magnitude masks 
%for the target speech and interferer 
%as in \cite{Wang2021Sequential}.
%We observe that complex spectral mapping is the best choice for single DNN MISO output.
%It outperforms the neural beamforming used in \cite{ren2021neural}. 
%It also outperforms magnitude-mask based mfMCWF for the reason outlined in Section \ref{sec:mfmcwf}: complex spectral mapping allows to produce an estimate aligned to the ground truth signal. 
%Regarding the mfMCWF, using complex spectral mapping allows for 

\begin{table}[]
\scriptsize
\centering
\caption{Results of one-DNN systems on dev. set. Approaches marked with * use additional STOI loss and ASR-based Deep Feature loss.
}
\vspace{-0.2cm}
\label{tab:singeDNN}
\setlength{\tabcolsep}{2.5pt}
\begin{tabular}{cccc}
\toprule

Approaches & WER (\%) & STOI & Task1 Metric  \\
 \midrule
Challenge Baseline \cite{ren2021neural} & 25.0 & 0.870 & 0.810 \\
\midrule

FasNet* \cite{luo2019fasnet} & 18.2 & 0.874 & 0.846 \\
Conv-TasNet \cite{luo2018convtasnet} MVDR* & 5.56 & 0.821 & 0.883 \\
DCCRN* \cite{hu2020dccrn} & 18.8 & 0.907 & 0.860 \\
Demucs v2* \cite{defossez2019demucs} & 26.3 &  0.851 & 0.794 \\
Demucs v3* \cite{defossez2021hybrid} & 15.3 & 0.874 &  0.860 \\

\midrule

DNN$_1$ & \textbf{3.90} & \textbf{0.964} & \textbf{0.963} \\ % 3.89999 & 0.964313 &  \\

\bottomrule
%\vspace{1ex}
%{\raggedright \footnotesize{} \par}
\end{tabular}
%\begin{tablenotes}
%      \small
%      \item \footnotesize{*Use additional STOI loss and ASR-based Deep Feature loss.}
%    \end{tablenotes}
\vspace{-0.4cm}
\end{table}

\begin{table}[]
\scriptsize
\centering
\caption{Comparison of various beamforming setup on dev. set.}
\vspace{-0.2cm}
\label{tab:mfmcwf}
\setlength{\tabcolsep}{2.5pt}
\begin{tabular}{lccccc}
\toprule

Approaches & $l$ & $r$ & WER (\%) & STOI & Task1 Metric  \\
\midrule

DNN$_1$ & - & - & 3.90 & 0.964 & 0.963 \\
DNN$_1$+mfMCWF & 0 & 0 & 6.98 & 0.917 & 0.923 \\
DNN$_1$+mfMCWF & 7 & 0 & 3.42 & 0.966 & 0.966 \\
DNN$_1$+mfMCWF& 6 & 1 & 3.13 & 0.974 & 0.971 \\
DNN$_1$+mfMCWF & 5 & 2 & 3.09 & 0.974 & 0.972 \\
DNN$_1$+mfMCWF & 4 & 3 & \textbf{3.04} & \textbf{0.975} & \textbf{0.972} \\

\midrule

Magnitude-mask based mfMCWF \cite{Wang2021Sequential} & 4 & 3 & 4.82 & 0.959 & 0.955 \\

\bottomrule
\end{tabular}
\vspace{-0.2cm}
\end{table}

%We report the results obtained on the L3DAS22 by our proposed method as well as other state-of-the-art multi-cna

%\subsection{Mask-Based Beamforming: STOI versus WER Trade-Off}

Table~\ref{tab:singeDNN} compares the challenge metrics obtained by the different models introduced in Section \ref{sec:benchmark}. % which have been explored during the challenge period. 
For these models we used additional losses related to the challenge metrics: namely the STOI loss and a \ac{DFL} derived from the Wav2Vec2 \ac{ASR} back-end used by the challenge to compute the WER scores. 
In detail we used as DFL the log-\ac{MSE} between the Wav2Vec2 final-layer activations when it is fed the enhanced signal versus when it is fed the oracle target speech signal. 
% In Fig. \ref{fig:STOI_WER}, we compare the trade-off between STOI and ASR performance based on Conv-TasNet approach. The 
% We found the trends of adding loss to approcah the clean speech improves the STOI more than the WER, such as adding STOI loss and post-filter loss. In contrast, adding DFL and estimating masks improves WER more than STOI, since the DFL is comparing the speech for the recognition task and masks estimation reduces the distortion for the enhanced speech.
Despite the proposed model is trained in a back-end agnostic way, i.e. without using DFL and STOI related losses, it significantly outperforms the other models which instead rely on additional loss terms associated with the particular challenge task. 
In addition, a noticeable trend is that the models employing complex spectral mapping (DNN$_1$, DCCRN and Demucs v2 and v3) consistently obtain higher STOI than Conv-TasNet MVDR, which is based on mask-based beamforming. 
%This aligns with our findings in Table~\ref{tab:dnn1_ablation}.
The models that rely on complex spectral mapping, being unconstrained, are capable of producing an aligned estimate with respect to the true oracle signal, leading to inherently higher STOI. In contrast, mask-based beamforming methods, as explained in Section \ref{mfmcwf}, produce an estimate that is constrained to be aligned with the input mixture signals. 
%We can speculate that this alignment is possible in this particular challenge because the acoustical environment is consistent between train, development and evaluation. This opens up an interesting research direction: whether or not this will also be possible when the model has to generalize to many different acoustic scenarios. 

\begin{table}[]
\scriptsize
\centering
\caption{Results of two-DNN systems on dev. set.}
\vspace{-0.2cm}
\label{tab:twoDNN}
\setlength{\tabcolsep}{2.5pt}
\begin{tabular}{lccccc}
\toprule

Approaches & $l$ & $r$ & WER (\%) & STOI & Task1 Metric  \\
\midrule
Challenge Baseline \cite{ren2021neural} & - & - & 25.0 & 0.870 & 0.810 \\
\midrule

DNN$_1$ & - & - & 3.90 & 0.964 &  0.963  \\ % 3.89999 & 0.964313 &  \\
DNN$_1$+MVDR+DNN$_2$ & - & - & 3.62 & 0.970 & 0.968 \\
DNN$_1$+mfMCWF+DNN$_2$ & 0 & 0 & 3.36 & 0.971 & 0.969 \\
DNN$_1$+mfMCWF+DNN$_2$ & 7 & 0 & 2.63 & 0.978 & 0.976 \\
DNN$_1$+mfMCWF+DNN$_2$ & 6 & 1 & 2.36 & 0.982 & 0.979 \\
DNN$_1$+mfMCWF+DNN$_2$ & 5 & 2 & 2.53 & 0.982 & 0.978 \\
DNN$_1$+mfMCWF+DNN$_2$ & 4 & 3 & 2.35 & 0.983 & 0.980 \\ %& 2.34878 & 0.982629 & \\
DNN$_1$+(mfMCWF+DNN$_2$)$\times$2 & 4 & 3 & \textbf{2.14} & \textbf{0.986} & \textbf{0.982} \\ % 2.14253 & 0.986063 & 0.982  \\

\bottomrule
\end{tabular}
\vspace{-0.4cm}
\end{table}

In Table~\ref{tab:mfmcwf}, we first report the mfMCWF results by using DNN$_1$'s output to compute the beamformer.
We set $l$ and $r$ to different values.
We can see that mfMCWF consistently outperforms single-frame MCWF, which is the same as mfMCWF with $l=0$ and $r=0$.
The best performance is obtained by using a quasi-symmetrical configuration of $l=4$ past frames and $r=3$ future frames, and the resulting linear mfMCWF even obtains better scores than the non-linear DNN$_1$.
For comparison, we also report the result of the magnitude-mask based mfMCWF in \cite{Wang2021Sequential}.
In this latter model, we slightly modify the TCN-DenseUNet architecture, and train through the mask based mfMCWF and compute the loss in Eq.~(\ref{wav+mag}) based on the beamformed signal.
We found this training-through mechanism essential to make the mask-based mfMCWF work.
We tried using the DNN$_1$'s output to derive a magnitude mask and compute the beamformer (i.e. without training-through).
However, this led to severely degraded performance, because the DNN$_1$'s output is not time- and gain-aligned with the far-field mixture and hence it is not straightforward how to compute a valid magnitude mask.
Also, the mask based mfMCWF needs to designate one of the microphones as the reference, meaning that the resulting beamformed signal cannot be fully aligned with the dry source signal. For this reason, complex spectral mapping for mfMCWF computation leads to higher performance.
In Table~\ref{tab:twoDNN} we report the results of including DNN$_2$ into our system.
Clear improvement is obtained over DNN$_1$ and DNN$_1$+mfMCWF.
Run-time iterative estimation (up to two iterations), denoted as DNN$_1$+(mfMCWF+DNN$_2$)$\times$2, further improves the performance, at a cost of increased computational requirements.
%Regarding the choice of beamforming module, mfMCWF consistently outperforms MVDR and MCWF (same as mfMCWF with $l=0, r=0$). The best performance is obtained by using a quasi-symmetrical configuration of $l=4$ past frames and $r=3$ future frames.

\begin{table}[]
\scriptsize
\centering
\caption{Results of two-DNN systems on eval. set.}
\vspace{-0.2cm}
\label{tab:eval_set}
\setlength{\tabcolsep}{2.5pt}
\begin{tabular}{lccccc}
\toprule

Approaches & $l$ & $r$ & WER (\%) & STOI & Task1 Metric  \\
  
\midrule

DNN$_1$ & - & - & 3.73 & 0.964 & 0.964 \\
DNN$_1$+mfMCWF+DNN$_2$ & 0 & 0 & 3.15 & 0.971 & 0.970 \\
DNN$_1$+mfMCWF+DNN$_2$ & 7 & 0 & 2.28 & 0.978 & 0.978 \\
DNN$_1$+mfMCWF+DNN$_2$ & 4 & 3 & 2.11 & 0.983 & 0.981 \\ 
DNN$_1$+(mfMCWF+DNN$_2$)$\times$2 & 4 & 3 & \textbf{1.89} & \textbf{0.987} & \textbf{0.984} \\

\midrule

Challenge baseline \cite{ren2021neural} & - & - & 21.2 & 0.878 & 0.833 \\ 
Runner-up system (BaiduSpeech) & - & - & 2.50 & 0.975 & 0.975 \\ 
		
\bottomrule
\end{tabular}
\vspace{-0.2cm}
\end{table}

\begin{table}[]
\scriptsize
\centering
\caption{Results of using one Ambisonic microphone on dev. set.}
\vspace{-0.2cm}
\label{tab:twoDNN1arr}
\setlength{\tabcolsep}{2.5pt}
\begin{tabular}{lccccc}
\toprule

Approaches & $l$ & $r$ & WER (\%) & STOI & Task1 Metric  \\
\midrule
%Challenge Baseline \cite{ren2021neural} & - & - &  &  &  \\
%\midrule
DNN$_1$ & - & - & 4.11 & 0.958 &  0.958  \\ % 3.89999 & 0.964313 &  \\
DNN$_1$+mfMCWF+DNN$_2$ & 12 & 3 & 2.45 & 0.980 & 0.978 \\ %& 2.34878 & 0.982629 & \\
DNN$_1$+(mfMCWF+DNN$_2$)$\times$2 & 12 & 3 & \textbf{2.49} & \textbf{0.982} & \textbf{0.979} \\ % 2.14253 & 0.986063 & 0.982  \\

\bottomrule
\end{tabular}
\vspace{-0.4cm}
\end{table}

In Table~\ref{tab:eval_set} we report the results obtained on the challenge evaluation set by a subset of configurations explored in Table~\ref{tab:twoDNN}. 
We notice that the results between the development set and evaluation set are consistent.
Our proposed approach ranked first among all the submissions\footnote{See https://www.l3das.com/icassp2022/results.html for the full ranking.} to the L3DAS22 Task 1 speech enhancement challenge and shows a remarkable improvement over the baseline system and a significant improvement over the runner-up system.

In Table~\ref{tab:twoDNN1arr} we additionally provide the results obtained by only using the first ambisonic microphone
%(in this case $P=4$)
for testing.
The signals at both ambisonic microphones are used for training, and this doubles the number of training examples.
The number of filter taps for mfMCWF is increased from 8 to 16.
The results on the development set are close to the ones obtained by using both ambisonic microphones for training and testing (compare the last rows of Table~\ref{tab:twoDNN1arr} and \ref{tab:twoDNN}).

%reports the results of using two DNNs.
%Comparing entry 0 with the others, we observe clear improvements by using beamforming and DNN$_2$ over DNN$_1$.
%Comparing entry 1 and 2 with the others, we notice that mfMCWF produces better performance than MVDR and single-frame MCWF.
%The results in entries 3 and 4 show that using future frames for mfMCWF leads to further improvement over only using the current and past frames.
%Finally, running DNN$_2$ for a second pass as in DNN$_1$+(mfMCWF+DNN$_2$)$\times$2, yields another slight improvement. No further improvement was observed with a third pass.

%We use the $\hat{S}_q^{(3)}$ in Sec~\ref{Sec:System Description} as the final enhancement result. 

%The score of the enhancement results is in Table \ref{tab:Dev results}, where the enhancement of the development set from our system achieves $0.982$ in the overall Task1 Metric.

% \begin{figure}[]
%  \centering
%  \includegraphics[width=\linewidth]{figs/WER-STOI.pdf}
%  \caption{The STOI-(1-WER) graph for Conv-Tasnet with MVDR approaches.} 
%  \label{fig:STOI_WER}
% \end{figure}

\vspace{-0.2cm}
\section{Conclusions}
\vspace{-0.2cm}

In this paper we have described our submission to the L3DAS22 Task 1 challenge. Our proposed iNeuBe framework relies on an iterative pipeline of linear beamforming and DNN-based complex spectral mapping. 
In our method, two DNNs are employed in a MISO configuration and use complex spectral mapping to estimate the target speech signal. The first DNN output is used to drive an mfMCWF, and a second DNN, taking the outputs of the first DNN and the mfMCWF as additional input features, is used to further refine the estimated target speech signal from the first DNN. The second DNN and linear beamforming can be run iteratively and we show that up to the second iterations there are noticeable improvements, especially regarding WER.

Compared to previous work, we propose here the use of mfMCWF
and show that computing mfMCWF weights using DNN-based complex spectral mapping output can have significant advantages in the challenge scenario. Our proposed method ranked first in the L3DAS22 challenge, significantly outperforming the baseline and the second-best system. 
As additional contributions we also performed several ablation studies weighting different configurations and the contribution of each block in the iNeuBe framework. 

Finally, we also compared our proposed approach with multiple state-of-the-art models and showed that it can achieve remarkably better challenge metrics, with both lower WER and higher STOI, even when the competing models are trained with back-end task aware losses. 

%The framework allows for run-time iterative processing which we show can additionally improve performance. 

%We have investigated multiple time- and complex-domain models for the 3D speech enhancement task of the L3DAS22 challenge.
%Evaluation results suggest that multi-microphone complex spectral mapping produces clearly better performance than other state-of-the-art time-domain approaches.
%Among them, TCN-DenseUNet based multi-microphone complex spectral mapping produces the best performance.
%Our proposed mfMCWF leads to clear gains over single-frame MCWF.
%Using a second DNN for post-filtering leads to further improvement. 

\vspace{-0.2cm}
\section{Acknowledgements}
\vspace{-0.2cm}

S. Cornell was partially supported by Marche Region within the funded project ``Miracle'' POR MARCHE FESR 2014-2020.
Z.-Q. Wang used the Extreme Science and Engineering Discovery Environment ~\cite{xsede}, supported by NSF grant number ACI-1548562, and the Bridges system~\cite{nystrom2015bridges},
supported by NSF award number ACI-1445606, at the Pittsburgh Supercomputing Center.

\bibliographystyle{IEEEtran}
\bibliography{refs}

\begin{thebibliography}{10}
\providecommand{\url}[1]{#1}
\csname url@samestyle\endcsname
\providecommand{\newblock}{\relax}
\providecommand{\bibinfo}[2]{#2}
\providecommand{\BIBentrySTDinterwordspacing}{\spaceskip=0pt\relax}
\providecommand{\BIBentryALTinterwordstretchfactor}{4}
\providecommand{\BIBentryALTinterwordspacing}{\spaceskip=\fontdimen2\font plus
\BIBentryALTinterwordstretchfactor\fontdimen3\font minus
  \fontdimen4\font\relax}
\providecommand{\BIBforeignlanguage}[2]{{%
\expandafter\ifx\csname l@#1\endcsname\relax
\typeout{** WARNING: IEEEtran.bst: No hyphenation pattern has been}%
\typeout{** loaded for the language `#1'. Using the pattern for}%
\typeout{** the default language instead.}%
\else
\language=\csname l@#1\endcsname
\fi
#2}}
\providecommand{\BIBdecl}{\relax}
\BIBdecl

\bibitem{vincent2018audio}
E.~Vincent, T.~Virtanen, and S.~Gannot, \emph{Audio source separation and
  speech enhancement}.\hskip 1em plus 0.5em minus 0.4em\relax John Wiley \&
  Sons, 2018.

\bibitem{WDL2018}
D.~Wang and J.~Chen, ``Supervised speech separation based on deep learning: An
  overview,'' \emph{IEEE/ACM Trans. Audio, Speech, Lang. Process.}, vol.~26,
  no.~10, pp. 1702--1726, 2018.

\bibitem{heymann_blstm}
J.~Heymann, L.~Drude, A.~Chinaev, and R.~Haeb-Umbach, ``{BLSTM} supported {GEV}
  beamformer front-end for the 3rd {CHiME} challenge,'' in \emph{Proc. ASRU},
  2015.

\bibitem{Boeddeker_practical}
C.~Boeddeker, H.~Erdogan, T.~Yoshioka, and R.~Haeb-Umbach, ``Exploring
  practical aspects of neural mask-based beamforming for far-field speech
  recognition,'' in \emph{Proc. ICASSP}, 2018.

\bibitem{erdogan2016improved}
H.~Erdogan, J.~R.~Hershey, S.~Watanabe \emph{et~al.}, ``Improved {MVDR}
  beamforming using single-channel mask prediction networks.'' in \emph{Proc.
  Interspeech}, 2016.

\bibitem{ochiai_unified_2017}
T.~Ochiai, S.~Watanabe, T.~Hori \emph{et~al.}, ``Unified architecture for
  multichannel end-to-end speech recognition with neural beamforming,''
  \emph{IEEE Journal of Selected Topics in Signal Processing}, vol.~11, no.~8,
  pp. 1274--1288, 2017.

\bibitem{Wang2021Sequential}
Z.-Q. Wang, H.~Erdogan, S.~Wisdom \emph{et~al.}, ``Sequential multi-frame
  neural beamforming for speech separation and enhancement,'' in \emph{Proc.
  SLT}, 2021.

\bibitem{luo2019fasnet}
Y.~Luo, C.~Han, N.~Mesgarani \emph{et~al.}, ``{FasNet}: Low-latency adaptive
  beamforming for multi-microphone audio processing,'' in \emph{Proc. ASRU},
  2019.

\bibitem{ren2021neural}
X.~Ren, L.~Chen, X.~Zheng \emph{et~al.}, ``A neural beamforming network for
  {B-Format} {3D} speech enhancement and recognition,'' in \emph{Proc. MLSP},
  2021.

\bibitem{tolooshams2020channel}
B.~Tolooshams, R.~Giri, A.~H. Song, U.~Isik, and A.~Krishnaswamy,
  ``Channel-attention dense {U-Net} for multichannel speech enhancement,'' in
  \emph{Proc. ICASSP}, 2020.

\bibitem{liu2020multichannel}
C.-L. Liu, S.-W. Fu, Y.-J. Li \emph{et~al.}, ``Multichannel speech enhancement
  by raw waveform-mapping using fully convolutional networks,'' \emph{IEEE/ACM
  Trans. Audio, Speech, Lang. Process.}, vol.~28, pp. 1888--1900, 2020.

\bibitem{Wang2021MCCSM}
Z.-Q. Wang, P.~Wang, and D.~Wang, ``Multi-microphone complex spectral mapping
  for utterance-wise and continuous speaker separation,'' \emph{IEEE/ACM Trans.
  Audio, Speech, Lang. Process.}, vol.~29, pp. 2001--2014, 2021.

\bibitem{Kinoshita2017}
K.~Kinoshita, M.~Delcroix, H.~Kwon \emph{et~al.}, ``Neural network-based
  spectrum estimation for online {WPE} dereverberation,'' in \emph{Proc.
  Interspeech}, 2017.

\bibitem{heymann2016neural}
J.~Heymann, L.~Drude, and R.~Haeb-Umbach, ``Neural network based spectral mask
  estimation for acoustic beamforming,'' in \emph{Proc. ICASSP}, 2016.

\bibitem{le2019sdr}
J.~Le~Roux, S.~Wisdom, H.~Erdogan, and J.~R. Hershey, ``Sdr--half-baked or well
  done?'' in \emph{Proc. ICASSP}, 2019.

\bibitem{taal2011algorithm}
C.~H. Taal, R.~C. Hendriks, R.~Heusdens, and J.~Jensen, ``An algorithm for
  intelligibility prediction of time--frequency weighted noisy speech,''
  \emph{IEEE Trans. Audio, Speech, Lang. Process.}, vol.~19, no.~7, pp.
  2125--2136, 2011.

\bibitem{iwamoto2022bad}
K.~Iwamoto, T.~Ochiai, M.~Delcroix \emph{et~al.}, ``How bad are artifacts?:
  Analyzing the impact of speech enhancement errors on {ASR},'' \emph{arXiv
  preprint arXiv:2201.06685}, 2022.

\bibitem{zhang2021closing}
W.~Zhang, J.~Shi \emph{et~al.}, ``Closing the gap between time-domain
  multi-channel speech enhancement on real and simulation conditions,'' in
  \emph{Proc. WASPAA}, 2021, pp. 146--150.

\bibitem{bagchi2018spectral}
D.~Bagchi, P.~Plantinga, A.~Stiff, and E.~Fosler-Lussier, ``Spectral feature
  mapping with mimic loss for robust speech recognition,'' in \emph{Proc.
  ICASSP}, 2018, pp. 5609--5613.

\bibitem{li2021espnet}
C.~Li, J.~Shi, W.~Zhang \emph{et~al.}, ``{ESPnet-SE}: end-to-end speech
  enhancement and separation toolkit designed for {ASR} integration,'' in
  \emph{Proc. SLT}, 2021.

\bibitem{guizzo2021l3das22}
E.~Guizzo, C.~Marinoni, M.~Pennese \emph{et~al.}, ``{L3DAS22} challenge:
  Learning {3D} audio sources in a real office environment,'' in \emph{Proc.
  ICASSP}, 2022.

\bibitem{Wang2021FCP}
Z.-Q. Wang, G.~Wichern, and J.~{Le Roux}, ``Convolutive prediction for monaural
  speech dereverberation and noisy-reverberant speaker separation,''
  \emph{IEEE/ACM Trans. Audio, Speech, Lang. Process.}, vol.~29, pp.
  3476--3490, 2021.

\bibitem{Wang2021LowDistortion}
------, ``Leveraging low-distortion target estimates for improved speech
  enhancement,'' \emph{arXiv preprint arXiv:2110.00570}, 2021.

\bibitem{nakatani2019unified}
T.~Nakatani and K.~Kinoshita, ``A unified convolutional beamformer for
  simultaneous denoising and dereverberation,'' \emph{IEEE Signal Processing
  Letters}, vol.~26, no.~6, pp. 903--907, 2019.

\bibitem{zhang2020multi}
Z.~Zhang, Y.~Xu \emph{et~al.}, ``{Multi-Channel Multi-Frame ADL-MVDR for Target
  Speech Separation},'' \emph{IEEE/ACM Trans. Audio, Speech, Lang. Process.},
  vol.~29, pp. 3526--3540, 2021.

\bibitem{panayotov2015librispeech}
V.~Panayotov, G.~Chen, D.~Povey, and S.~Khudanpur, ``Librispeech: an {ASR}
  corpus based on public domain audio books,'' in \emph{Proc. ICASSP}, 2015.

\bibitem{fonseca2020fsd50k}
E.~Fonseca, X.~Favory, J.~Pons \emph{et~al.}, ``{FSD50K}: An open dataset of
  human-labeled sound events,'' \emph{IEEE/ACM Trans. Audio, Speech, Lang.
  Process.}, 2021.

\bibitem{taal2010short}
C.~H. Taal, R.~C. Hendriks, R.~Heusdens, and J.~Jensen, ``A short-time
  objective intelligibility measure for time-frequency weighted noisy speech,''
  in \emph{Proc. ICASSP}, 2010.

\bibitem{baevski2020wav2vec}
A.~Baevski, H.~Zhou, A.~Mohamed, and M.~Auli, ``wav2vec 2.0: A framework for
  self-supervised learning of speech representations,'' \emph{arXiv preprint
  arXiv:2006.11477}, 2020.

\bibitem{Wang2020MCCSMDereverb}
Z.-Q. Wang and D.~Wang, ``Multi-microphone complex spectral mapping for speech
  dereverberation,'' in \emph{Proc. ICASSP}, 2020, pp. 486--490.

\bibitem{Tan2022}
K.~Tan, Z.-Q. Wang, and D.~Wang, ``Neural spectrospatial filtering,''
  \emph{IEEE/ACM Trans. Audio, Speech, Lang. Process.}, vol.~30, pp. 605--621,
  2022.

\bibitem{Wang2021compensation}
Z.-Q. Wang, G.~Wichern, and J.~Le~Roux, ``On the compensation between magnitude
  and phase in speech separation,'' \emph{IEEE Signal Process. Lett.}, vol.~28,
  pp. 2018--2022, 2021.

\bibitem{hu2020dccrn}
Y.~Hu, Y.~Liu, S.~Lv \emph{et~al.}, ``{DCCRN}: Deep complex convolution
  recurrent network for phase-aware speech enhancement,'' \emph{Proc.
  Interspeech}, 2020.

\bibitem{defossez2019demucs}
A.~D{\'e}fossez, N.~Usunier, L.~Bottou, and F.~Bach, ``Demucs: Deep extractor
  for music sources with extra unlabeled data remixed,'' \emph{arXiv preprint
  arXiv:1909.01174}, 2019.

\bibitem{cornell2021fbanks}
S.~Cornell, M.~Pariente, F.~Grondin, and S.~Squartini, ``Learning filterbanks
  for end-to-end acoustic beamforming,'' \emph{arXiv e-prints}, pp.
  arXiv--2111, 2021.

\bibitem{luo2018convtasnet}
Y.~Luo and N.~Mesgarani, ``{Conv-TasNet}: Surpassing ideal time–frequency
  magnitude masking for speech separation,'' \emph{IEEE/ACM Trans. Audio,
  Speech, Lang. Process.}, vol.~27, no.~8, p. 1256–1266, 2019.

\bibitem{luo2020end}
Y.~Luo, Z.~Chen, N.~Mesgarani, and T.~Yoshioka, ``End-to-end microphone
  permutation and number invariant multi-channel speech separation,'' in
  \emph{Proc. ICASSP}, 2020.

\bibitem{defossez2021hybrid}
A.~D{\'e}fossez, ``Hybrid spectrogram and waveform source separation,''
  \emph{Proc. ISMIR}, 2021.

\bibitem{xsede}
J.~Towns, T.~Cockerill, M.~Dahan \emph{et~al.}, ``{XSEDE}: Accelerating
  scientific discovery,'' \emph{Computing in Science \& Engineering}, vol.~16,
  no.~5, pp. 62--74, 2014.

\bibitem{nystrom2015bridges}
N.~A. Nystrom, M.~J. Levine, R.~Z. Roskies, and J.~R. Scott, ``Bridges: a
  uniquely flexible {HPC} resource for new communities and data analytics,'' in
  \emph{Proc. XSEDE}, 2015, pp. 1--8.

\end{thebibliography}

\end{document}